\documentclass[12pt]{article}
\usepackage{amsmath}
\usepackage{amssymb}
\usepackage{mathtext}
\usepackage{graphicx}

\begin{document}

\section*{KOVALEVSKAYA TOP AND GENERALIZATIONS OF
INTEGRABLE SYSTEMS\footnotemark}
\markboth{KOVALEVSKAYA TOP AND GENERALIZATIONS OF
INTEGRABLE SYSTEMS}{KOVALEVSKAYA TOP AND GENERALIZATIONS OF
INTEGRABLE SYSTEMS}

\footnotetext{REGULAR AND
CHAOTIC DYNAMICS V.6, No. 1, 2001

{\it Received December 12, 2000

AMS MSC 70E17, 70E40}}

\begin{centering}
A.\,V.\,BORISOV\\
Faculty of Mechanics and Mathematics,\\
Department of Theoretical Mechanics
Moscow State University\\
Vorob'ievy gory, 119899 Moscow, Russia\\
E-mail: borisov@uni.udm.ru\medskip\\
I.\,S.\,MAMAEV\\
Laboratory of Dynamical Chaos and Nonlinearity,\\
Udmurt State University\\
Universitetskaya, 1, Izhevsk, Russia, 426034\\
E-mail: mamaev@uni.udm.ru\medskip\\
A.\,G.\,KHOLMSKAYA\\
Udmurt State University\\
Universitetskaya, 1, Izhevsk, Russia, 426034\\
\end{centering}

\begin{abstract}
Generalizations of the Kovalevskaya, Chaplygin,
Goryachev--Chaplygin and Bogoyavlensky systems on a bundle
are considered in this paper. Moreover, a method of introduction of separating
variables and action--angle variables is described. Another integration
method for the Kovalevskaya top on the bundle is found. This method uses
a coordinate transformation that reduces the Kovalevskaya system to
the Neumann system. The Kolosov analogy is considered.
A generalization of a recent Gaffet system to the bundle of Poisson brackets
is obtained at the end of the paper.
\end{abstract}

{\it Dedicated to the $150$-th anniversary of
S.\,V.\,Kovalevskaya}\bigskip

It was the 150-th anniversary on January 15, 2000 of Sofia Vasilievna Kovalevskaya,
an outstanding women-mathematician, whose works made a great contribution
into development of the world science.

Her literature works also were of great importance for
the Russian culture where she developed progressive and even revolutionary
ideas.
\goodbreak

The most important results were obtained by S.\,V.\,Kovalevskaya
in the theory of differential equations and in rigid body dynamics.
S.\,V.\,Kovalevskaya stated and proved the existence and uniqueness theorem
 for solutions of partial differential
equations~\cite{kovdif}
(Cauchy-Kovalevskaya theorem). She also found a new method of
integrability analysis (the so-called Painlev\'e--Kovalevskaya test).
The substance of this test is as follows.

Let a system of differential equations
\begin{equation}
\label{intro1}
z_i=f_i(z_1,\,\ldots,\,z_n),\quad 1\le i\le n
\end{equation}
be invariant with respect to the similarity transformations
$$
t\to t/\alpha,\quad z_i\to \alpha^{g_i}z_i,\quad 1\le i\le n.
$$
Let us construct a matrix (the Kovalevskaya matrix)
$$
K_{ij}= \frac{\partial f_i}{\partial z_j}(c) +g_j\delta_{ij},
$$
where the constants $c_i$ satisfy the algebraic system of equations
$f_i(c_1,\,\ldots,\,c_n)=-g_ic_i$ ($1\le i\le n$).
Kovalevskaya required the conditions of integrality and nonnegativity
of eigenvalues of~$K$ for the uniqueness of
a general solution of~(\ref{intro1}) on a complex time plane.
Lyapunov~\cite{lyap} continued these investigations of Kovalevskaya and
related the meromorphy of a general solution with properties of
the variational equations system. The Lyapunov method was further developed
by S.\,L.\,Ziglin, who proved strict results concerning nonexistence of
additional single-valued integrals.

The works on rotation of a heavy rigid body about a fixed point
are the most famous~\cite{kov} (1888). In these works a new integrable case
of the Euler--Poisson equations  was found and its solution in quadratures
was obtained. The French Academy of Sciences highly appreciated
this S.\,V.\,Kovalevskaya result and awarded her with the Prix Bordin in 1888.
Let us note that the Academy of Sciences had twice announced the
prize contest on investigations on this subject, however the prize was not
awarded. S.\,V.\,Kovalevskaya was awarded with a prize of the Sweden Academy of
Sciences for the second memoir on the problem of rotation
of a rigid body in spring~1889. A lot of  questions concerning the motion of
the Kovalevskaya top are still open.
\medskip

\centerline{\rule{2cm}{0.4pt}}\medskip

All integrable cases under consideration are well known in rigid body dynamics.
However, not for all of them the generalizations from the algebra $e(3)$
to the bundle of brackets  $\mathcal{L}_x$
\begin{equation}
\label{KOV1separ}
\{M_i,M_j\}=-\varepsilon_{ijk}M_k,\quad
\{M_i,\gamma_j\}=-\varepsilon_{ijk}\gamma_k,\quad
\{\gamma_i,\gamma_j\}=-x\varepsilon_{ijk}M_k,
\end{equation}
have been investigated yet.
Let us note that  $\mathcal{L}_x$ possesses two quadratic Casimir functions
\begin{equation}
\label{KOV2separ}
\begin{aligned}
l&=({\boldsymbol M}, {\boldsymbol \gamma}),\\
c&=x({\boldsymbol M}, {\boldsymbol M}) + ({\boldsymbol \gamma}, {\boldsymbol \gamma}).
\end{aligned}
\end{equation}

These problems naturally appear for a system which describes dynamics
of a rigid body in an ideal fluid, connected bodies, motion of a body with
elliptic hollow filled in with a liquid, etc. Generalizations of known
integrable cases to the algebra $so(4)$ and separating variables for
several of them can be found in~\cite{Bogoyav}.

\section{The Kovalevskaya Top and its Generalizations}

\subsection{Integrals of the Kovalevskaya Case and its Generalizations}

S.\,V.\,Kovalevskaya found a new general integrable case of the
Euler-Poisson equations in 1888~\cite{kov} basing on pure mathematical
considerations. She was developing ideas of K.\,Weierstrass, P.\,Painlev\'e and
H.\,Poincar\'e concern the investigation of analytical extension of solutions of
a system of ordinary differential equations to the complex time plane.
Kovalevskaya assumed that a general solution does not possess other
singularities except for poles in the complex plane in integrable cases.
That enabled to find existence conditions for the additional integral.
Moreover, S.\,V.\,Kovalevskaya found a quite nonobvious system of variables.
The motion equations expressed in these variables are of the Abel-Jacobi
type.
She also obtained an explicit solution in theta-functions.

In the Kovalevskaya case the body is symmetric:
$I_1=I_2=2I_3$, where $I_i$ ($i=1,2,3$) are principal central
moments of inertia, the centre of mass is situated in the equatorial plane of
the inertia ellipsoid. The Hamiltonian has the form
\begin{equation}
\label{KOV3separ}
H=\frac{1}{2}(M_1^2+M_2^2+2M_3^2)+\gamma_1.
\end{equation}

The additional integral found by Kovalevskaya can be represented as follows:
\begin{equation}
\label{Def8}
F_3=\left(\frac{M_1^2-M_2^2}{2}-\boldsymbol\gamma_1\right)^2+
\left(M_1 M_2-\boldsymbol\gamma_2\right)^2
\end{equation}
(without loss of generality it is assumed that the
radius-vector of the centre of mass is ${\boldsymbol r}=(1,0,0)$, weight equals unit).
Let us note that the integral $F_3$ does not have an evident symmetry origin.

The Kovalevskaya case (on $e(3)$) can be generalized to the bundle of
brackets $\mathcal{L}_x$~(1).

Thus, the additional integral (the Kovalevskaya integral) will have the form
\begin{equation}
\label{KOV4}
\begin{array}{c}
k^2=k_1^2 + k_2^2,\\[8pt]
k_1 = M_1^2 - M_2^2 - 2\gamma_1 + x,\quad
k_2 = 2M_1M_2-2\gamma_2.
\end{array}
\end{equation}

Generalizations of integrable cases can be associated with
introduction of preserving integrability additional terms in the Hamiltonian
without the change of structure of the  Poisson brackets algebra
(for example, introduction of a gyrostat or singular terms).
Generalizations can be also associated with
the change of the algebra structure without change of the Hamiltonian
(for example, generalization to the bundle of brackets $\mathcal{L}_x$),
with a simultaneous change of the Hamiltonian and Poisson brackets,
with a generalization to the case of several force fields.

The motion equations can be generalized if one introduces a constant
gyrostatic moment induced by a balanced rotor which is fixed in
a rigid body and rotates with a constant angular velocity. An
analogous moment occurs in motion of a rigid body with multiply connected
 (that admits a possibility of appearance of nonzero circulation)
hollows containing an ideal incompressible liquid~\cite{zhuk}.
In this case the algebra and motion equations do not change. However,
a linear in momenta term appears in the Hamiltonian.

A gyrostat in the Kovalevskaya top was introduced by Yehia (1987) and Komarov
(1987):
$$
\begin{aligned}
H =&\frac {1}{2}\biggl(M_1^2+M_2^2+2\Bigl(M_3-\frac{\lambda}{2}\Bigr)^2\biggr)+r_1\gamma_1, \\
F ={}&(M_1^2-M_2^2-2r_1\gamma_1)^2+(2M_1M_2-2r_1\gamma_2)^2+ \\[2pt]
{}&+ 4\lambda(M_3-\lambda)(M_1^2+M_2^2)-8r_1\lambda M_1\gamma_3.
\end{aligned}
$$

In~\cite{goriachev} D.\,N.\,Goryachev suggested a generalization of the
Kovalevskaya case at the zero area constant $(\boldsymbol M,\,\boldsymbol \gamma)=0$
\begin{equation}
\label{gen3}
\begin{gathered}
\begin{aligned}
H={}&\frac{1}{2}(M_1^2 + M_2^2 + 2M_3^2)+\frac{a}{\gamma_3^2}+2b_1\gamma_1\gamma_2+b_2(\gamma_2^2-\gamma_1^2)+
r_1\gamma_1+r_2\gamma_2,\\
F={}&4\left(M_1M_2-2a\frac{\gamma_1\gamma_2}{\gamma_3^2}+b_1\gamma_3^2-r_1\gamma_2-
r_2\gamma_1\right)^2+ \\
{}&+ \left(M_1^2-M_2^2-2a\frac{(\gamma_1^2-\gamma_2^2)}{\gamma_3^2}-2b_2\gamma_3^2-
2r_1\gamma_1+2r_2\gamma_2\right)^2,\\
\end{aligned}\\
a,\; b_1,\; b_2,\; r_1,\; r_2=const.
\end{gathered}
\end{equation}

A physical sense of the singular term can be interpreted in quantum mechanics
and with the help of reduction with symmetry suggested in~\cite{brm}
for a variant of rigid body motion in superposition of two
homogeneous fields.  Let us note that at $a=0$ the integrable generalization
under consideration was indicated by  S.\,A.\,Chaplygin
in his work~\cite{chapl}, dedicated to a new integrable case of the Kirchhoff equations.
D.\,N.\,Goryachev did not refer to~\cite{chapl}, although his
work~\cite{goriachev} appeared later.
\goodbreak

In the Yehia work~\cite{Yehia} a constant gyrostatic moment along the axis
of dynamical symmetry is added in
the Hamiltonian~(\ref{gen3}) at the zero value of the area integral
$$
\begin{gathered}
\begin{aligned}
H={}&\frac{1}{2}(M_1^2 + M_2^2 + 2(M_3+\lambda)^2)+\frac{a}{\gamma_3^2}+2b_1\gamma_1\gamma_2+b_2(\gamma_2^2-\gamma_1^2)+
r_1\gamma_1+r_2\gamma_2,\\
F={}&4\left(M_1M_2-2a\frac{\gamma_1\gamma_2}{\gamma_3^2}+b_1\gamma_3^2-r_1\gamma_2-
r_2\gamma_1\right)^2+ \\
{}&+ \left(M_1^2-M_2^2-2a\frac{(\gamma_1^2-\gamma_2^2)}{\gamma_3^2}-2b_2\gamma_3^2-
2r_1\gamma_1+2r_2\gamma_2\right)^2-\\
{}&-8\lambda(M_3+2\lambda)(M_1^2+M_2^2)-16a\lambda(M_3+2\lambda)\left(1+\frac{1}{\gamma_3^2}
\right)+ \\
{}&+16\lambda\gamma_3\Bigl(M_1( r_1+b_1\gamma_2-b_2\gamma_1)+M_2(r_2+b_1\gamma_1+
b_2\gamma_2)\Bigr).
\end{aligned}\\
a,\; b_1,\; b_2,\; r_1,\; r_2,\; \lambda=const.
\end{gathered}
$$

In~\cite{yahia} Yehia suggested another generalization of the Kovalevskaya case
with the help of the introduction of a singular (but another) term,
which has not obtained a clear physical interpretation yet
$$
\begin{gathered}
\begin{aligned}
H={}&\frac{1}{2}(M_1^2+M_2^2+2M_3^2)+r_1\gamma_1+r_2\gamma_2+
\frac{\varepsilon}{\sqrt{\gamma_1^2+\gamma_2^2}},\\
F={}&(M_1^2-M_2^2-2r_1\gamma_1+2r_2\gamma_2)^2+
4(M_1M_2-r_1\gamma_2-r_2\gamma_1)^2+
+ 4\frac{\varepsilon(M_1^2+M_2^2)}
{\sqrt{\gamma_1^2+\gamma_2^2}}+4\frac{\varepsilon^2}{\gamma_1^2+\gamma_2^2},\\
\end{aligned}\\
\varepsilon,\; r_1,\; r_2=const.
\end{gathered}
$$
This integral curiously gives a general integrable case.

Another generalization of the Kovalevskaya case is associated with introduction
of two different singular potentials in the Hamiltonian~\cite{Yehia}:
$$
\begin{gathered}
\begin{array}{c}
H=\frac{1}{2}(M_1^2+M_2^2+2M_3^2)+r_1\gamma_1+r_2\gamma_2+
\frac{\varepsilon}{\sqrt{\gamma_1^2+\gamma_2^2}}+\frac{a}{\gamma_3^2},\\
a,\; \varepsilon,\; r_1,\; r_2=const.
\end{array}
\end{gathered}
$$
A particular integral at the zero area constant has the form:
$$
\begin{gathered}
\begin{aligned}
F={}&\left(M_1^2-M_2^2-2r_1\gamma_1+2r_2\gamma_2-\frac{2a(\gamma_1^2-\gamma_2^2)}
 {\gamma_3^2}\right)^2+ \\
{}&+ 4\left(M_1M_2-r_1\gamma_2-r_2\gamma_1-\frac{2a\gamma_1\gamma_2}{\gamma_3^2}\right)^2+ \\
{}&+ 4\varepsilon\left(\frac{M_1^2+M_2^2}
{\sqrt{\gamma_1^2+\gamma_2^2}}+\frac{\varepsilon}{\gamma_1^2+\gamma_2^2}+
\frac{2a\sqrt{\gamma_1^2+\gamma_2^2}}{\gamma_3^2}\right).\\
\end{aligned}
\end{gathered}
$$

Let us note that the integral $F$ is invalid in~\cite{Yehia}, though
it is probably connected with a strange misprint.

There exist generalizations of the Kovalevskaya and Chaplygin
integrable cases which include gyrostatic terms on the entire bundle~$\mathcal{L}_x$.
The analogue of the area constant is also supposed to be zero in this case
$$
  (\boldsymbol M,\,\boldsymbol \gamma)=0.
$$
It is more convenient to represent the Hamiltonian as follows:
\begin{equation}
\label{m7.1}
\begin{aligned}
H  ={}&\frac {1}{2}\Bigl((1+xa_1)M_1^2+(1+xa_2)M_2^2+2\Bigl(M_3-\frac\lambda
2\Bigr)^2\Bigr)+\\
{}& + a_2 \gamma_1^2+a_1\gamma_2^2+\frac {1}{2}(a_1+a_2)\gamma_3^2+r_1\gamma_1+r_2\gamma_2.
\end{aligned}
\end{equation}
This form of the Hamiltonian at~$\lambda=0$ differs from the representation
indicated in~\cite{Bogoyav} on the Casimir function
$\frac {1}{2}(a_1+a_2)(x\boldsymbol M^2+\boldsymbol \gamma^2)$. The result regarding
integrability of the system for $\lambda\ne 0$ is new.

In this case the integral has the form
\begin{equation}
\label{m7.2}
\begin{aligned}
K = {}&k_1^2+\alpha_1\alpha_2k_1^2-\lambda\{k_1,\,k_2\}-4\lambda^2(M_1^2+M_2^2),\\
k_1 = {}&\alpha_1M_1^2-\alpha_2M_2^2-(a_1-a_2)\gamma_3^2-2(\gamma_1r_1+\gamma_2r_2)+x\frac{\alpha_1
r_1^2-\alpha_2r_2^2}{\alpha_1\alpha_2},\\
k_2  ={}&M_1M_2-2\frac{\alpha_1 r_1 \gamma_2+\alpha_2 r_2 \gamma_1}{\alpha_1\alpha_2}+
2x\frac{r_1r_2}{\alpha_1\alpha_2},\\
\{k_1,\,k_2\}={}&4M_3\Bigl(\alpha_1 M_1^2-\alpha_2M_2^2-x\frac{\alpha_1r_1^2+\alpha_2r_2^2}
{\alpha_1\alpha_2}\Bigr)-4\gamma_3((a_1-a_2)(M_1\gamma_1-M_2\gamma_2)-\\
-2(M_1r_1+M_2r_2)).\\
\end{aligned}
\end{equation}

{\bf Remark.}  {\it
The motion equations for~$K$ can be represented in the form
$$
  \dot K=\{K,\,H\}=(a_1-a_2)(\boldsymbol M,\,\boldsymbol \gamma)F(\boldsymbol M,\,\boldsymbol \gamma).
$$
Thus, this case becomes a general integrable case at~$a_1=a_2$
that corresponds to the gyrostatic generalization of the
Kovalevskaya case.}

\subsection{Separation of Variables on the Bundle~\boldmath{$\mathcal{L}_x$}}

Our reduction to the  Abel equations on the bundle of brackets~$\mathcal{L}_x$
uses arguments of G.\,K.\,Suslov, who suggested a new method of
integration of the Kovalevskaya case in his famous treatise~\cite{suslov}.

Let us set new variables
\begin{alignat*}{3}
z_1={}&M_1 + iM_2,{}&\quad z_2 ={}& M_1 - iM_2,\\
\zeta_1={}& k_1 + ik_2,{}&\quad \zeta_2 ={}& k_1 -ik_2,\\[6pt]
{}\span
\hspace{25mm}\zeta_1\zeta_2=k^2.
\end{alignat*}
Let us take advantage of the motion equations for $z_1,z_2$
$$
i\dot z_1 = M_3z_1-\gamma_3,\quad -i\dot z_2 = M_3z_2-\gamma_3
$$
and formulate  $\gamma_1,\gamma_2,\gamma_3,M_3$ as
\begin{equation}
\label{KOV5}
\begin{aligned}
\gamma_1 ={}& \frac{1}{4}\left( z_1^2+z_2^2 \right)
           - \frac{1}{4}\left( \zeta_1 + \zeta_2 \right) + \frac{x}{2},\\[8pt]
\gamma_2 ={}& -\frac{i}{4}\left( z_1^2 - z_2^2 \right)
           +\frac{i}{4}\left( \zeta_1 - \zeta_2 \right),\\[8pt]
\gamma_3 ={}& i\frac{\dot z_1z_2 + \dot z_2z_1}{z_1 - z_2},\quad
M_3 = i\frac{\dot z_1 + \dot z_2}{z_1 - z_2}.
\end{aligned}
\end{equation}

Then insert the obtained expressions into the integrals (2)
and the Hamiltonian (\ref{KOV3separ}). Solving the obtained equations,
one finds
\begin{equation}
\label{KOV6}
\begin{aligned}
\dot z_1^2 ={}& R_1 -\frac{\zeta_1}{4} \left(z_1-z_2\right)^2,\\[6pt]
\dot z_2^2 ={}& R_2 - \frac{\zeta_2}{4} \left(z_1-z_2\right)^2,\\[8pt]
\dot z_1 \dot z_2 ={}& -R - \frac{1}{4} \left(2h-x\right)\left(z_1-z_2\right)^2,
\end{aligned}
\end{equation}
where
\begin{equation}
\label{KOV7}
\begin{array}{c}
R = R(z_1,z_2) = \frac{1}{4} z_1^2z_2^2 - \frac{h}{2}\left(z_1^2 + z_2^2\right)
                + l\left(z_1+z_2\right)+\frac{k^2}{4}-c
                + {xh}-\frac{x^2}{4},\\[8pt]
R_1 = R(z_1,z_1),\qquad R_2 = R(z_2,z_2),
\end{array}
\end{equation}
here $h$ is a constant of the energy integral  (\ref{KOV3separ}).

We still need to exclude $\zeta_1,\zeta_2$ from the obtained equations
with the help of the Kovalevskaya integral~(\ref{KOV4})
$$
\left(R_1-\dot z_1^2 \right)\left(R_2-\dot z_2^2 \right) =
\frac{\zeta_1\zeta_2}{16}(z_1-z_2)^4=\frac{k^2}{16}(z_1-z_2)^4.
$$

Rearrange terms in the obtained relation and represent the latter in the form
\begin{equation}
\label{KOV8}
\begin{array}{l}
\left(\frac{\dot z_1}{\sqrt{R_1}} + \frac{\dot z_2}{\sqrt{R_2}} \right)^2 =
\left(\frac{\dot z_1\dot z_2}{\sqrt{R_1R_2}} + 1 \right)^2
-\frac{k^2(z_1-z_2)^4}{16R_1R_2}=f_1\\[10pt]
\left(\frac{\dot z_1}{\sqrt{R_1}} - \frac{\dot z_2}{\sqrt{R_2}} \right)^2 =
\left(\frac{\dot z_1\dot z_2}{\sqrt{R_1R_2}} - 1 \right)^2
-\frac{k^2(z_1-z_2)^4}{16R_1R_2}=f_2,
\end{array}
\end{equation}
where $\dot z_1\dot z_2$ are substituted from (\ref{KOV6}).

The last step is the introduction of the Kovalevskaya variables by formulae
\begin{equation}
\label{KOV9}
s_1=\frac{R-\sqrt{R_1R_2}}{2(z_1-z_2)^2},\qquad
s_2=\frac{R+\sqrt{R_1R_2}}{2(z_1-z_2)^2}.
\end{equation}
Let us express
$R=(s_1+s_2)(z_1-z_2)^2,\,\sqrt{R_1R_2}=(s_2-s_1)(z_1-z_2)^2$
from these relations
and insert them
in the right-hand sides of (\ref{KOV8}). We obtain
\begin{equation}
\label{KOV9_1}
\begin{array}{c}
f_1=\frac{f(s_1)}{(s_1-s_2)^2},\quad
f_2=\frac{f(s_2)}{(s_1-s_2)^2},\\[8pt]
f(s)=\left(2s+\frac{1}{4}(2h-x)+\frac{1}{4}k\right)
      \left(2s+\frac{1}{4}(2h-x)-\frac{1}{4}k\right).
\end{array}
\end{equation}

The relations for the left-hand sides of (\ref{KOV8}) define a transition to
curvilinear coordinates
\begin{equation}
\label{KOV10}
\frac{dz_1}{\sqrt{R_1}}+\frac{dz_2}{\sqrt{R_2}}=
\frac{ds_1}{\sqrt{\varphi(s_1)}},\qquad
\frac{dz_1}{\sqrt{R_1}}-\frac{dz_2}{\sqrt{R_2}}=
\frac{ds_2}{\sqrt{\varphi(s_2)}},
\end{equation}
where $\varphi(s)$ is a polynomial of the third degree,
$$
\varphi(s) = 4s^3 + 2hs^2
+\left(\frac{1}{16}(2h-x)^2-\frac{k^2}{16}+\frac{c}{4}\right) s+\frac{l^2}{16}.
$$

Inserting (\ref{KOV10}) into (\ref{KOV8}), we obtain
the motion equations in the variables $s_1,s_2$
\begin{equation}
\label{KOV10_1}
\dot s_1=\frac{\sqrt{f(s_1)\varphi(s_1)}}{s_1-s_2},\quad
\dot s_2=\frac{\sqrt{f(s_2)\varphi(s_2)}}{s_2-s_1}.
\end{equation}

{\footnotesize
The polynomial $\varphi(s)$ can be obtained with the help of standard methods
using reduction of elliptic integrals to the standard from.
Let us describe the Suslov method~\cite{suslov} here.

The Kovalevskaya variables are solutions of the  quadratic equation
\begin{equation}
\label{KOV11}
Q(z_1,z_2,s)=(z_1-z_2)^2s^2-Rs+G=0,
\end{equation}
where
$$
G=\frac{R^2-R_1R_2}{4(z_1-z_2)^2} =
-\frac{l}{8}z_1z_2(z_1+z_2)+
\frac{1}{64}\left((2h-x)^2-k^2+4c\right)(z_1+z_2)^2
-\frac{lh}{4}(z_1+z_2)+\frac{l^2}{4}.
$$
Let us calculate quadrates of derivatives
$\left(\frac{\partial Q}{\partial s}\right)^2,
 \left(\frac{\partial Q}{\partial z_1}\right)^2,
 \left(\frac{\partial Q}{\partial z_2}\right)^2$
and using (\ref{KOV11}) exclude $s,z_1,z_2$ from them
respectively. By a direct calculation we verify that
$$
\left(\frac{\partial Q}{\partial s}\right)^2=R_1R_2,\quad
\left(\frac{\partial Q}{\partial z_1}\right)^2=\varphi(s)R_2,\quad
\left(\frac{\partial Q}{\partial z_2}\right)^2=\varphi(s)R_1.
$$

Let us construct an exact differential of $Q$
$$
dQ=\frac{\partial Q}{\partial z_1}dz_1
  +\frac{\partial Q}{\partial z_2}dz_2
  +\frac{\partial Q}{\partial s}ds=0.
$$
Dividing by $\sqrt{\varphi(s)R_1R_2}$ and taking into account a possibility of
rooting with different signs, we obtain the relation~(\ref{KOV10}).
}

\subsection{The Haine-Horozov Transformation for the Kovalevskaya System}

Let us consider another method of integration of the Kovalevskaya system.
This method uses a coordinate transformation analogous to the
one indicated in~\cite{Haine}. The Kovalevskaya system reduces to the
Neumann problem on a motion of a point on $S^2$
with the help of this transformation. That allows to write down a Lax
pair for the Kovalevskaya top and implies a relation between integrals of
both systems.

Let us denote the motion integrals of the Kovalevskaya system
on  $\mathcal{L}_x$ as follows:
\begin{equation}
\begin{aligned}
\label{horoz3}
2H=\frac{1}{2}(M_1^2+M_2^2+2M_3^2) + 2\gamma_1=h,\\
c_1=(\boldsymbol M,\,\boldsymbol\gamma),\quad
c_2=x (\boldsymbol M,\,\boldsymbol M) + (\boldsymbol\gamma,\,\boldsymbol\gamma)-k^2,\\
k^2=\left(\frac{M_1^2-M_2^2}{4}-\gamma_1+x \right)^2+
\left( \frac{M_1M_2}{2} -\gamma_2 \right)^2
\end{aligned}
\end{equation}
and transform variables by formulae
\begin{equation}
\begin{aligned}
\label{horoz1}
\begin{array}{ll}
p_1=\frac{M_1^2+M_2^2+4}{4M_2}, &\quad l_1=\frac{4M_1\gamma_3-M_3(M_1^2+M_2^2-4)}{8M_2},\\
p_2=-i\frac{M_1}{M_2}, &\quad l_2=-i\frac{\gamma_3}{M_2},\\
p_3=i\frac{M_1^2+M_2^2-4}{4M_2}, &\quad l_3=i\frac{4M_1\gamma_3-M_3(M_1^2+M_2^2+4)}{8M_2}.
\end{array}
\end{aligned}
\end{equation}

Then after the coordinate transformation~(\ref{horoz1}) the motion equations
can be written as
\begin{equation}
\label{horoz4}
\dot{\boldsymbol  p}=\boldsymbol l\times\boldsymbol p, \quad \dot{\boldsymbol l}=
{\bf Q}{\boldsymbol p}\times\boldsymbol p+
\frac{x(x-h)}{4}(p_1+ip_3) \boldsymbol y,
\end{equation}

where the matrix ${\bf Q}$ has the form
\begin{equation}
\label{horoz2}
{\bf Q}=\frac{1}{4}\left(
\begin{array}{ccc}
c_2-1 & -ic_1 & i(1+c_2) \\
-ic_1 & -h & c_1 \\
i(1+c_2) & c_1 & 1-c_2
\end{array}
\right),
\end{equation}
  $h$, $c_1$, $c_2$, $k^2$
are determined by~(\ref{horoz3}), the vector $\boldsymbol y=(-ip_2,\,i(p_1+ip_3),\,p_2)$.
\goodbreak

The motion equations~(\ref{horoz4}) can be regarded as the system on
 $e(3)$ with the Hamiltonian
$$
(\boldsymbol l,\,\boldsymbol l)+({\bf Q}{\boldsymbol p},\,\boldsymbol p)+\frac{x(x-h)}{4}(p_1+ip_3)^2-x
=-\frac{h}{4},
$$
i.\,e. as the Neumann system. Thus, the Kovalevskaya system on
the bundle~(1) can be integrated by reduction to the
Neumann system. The integration of the latter is performed by standard methods.

Let us note that the Kovalevskaya system can not be  reduced
to the Neumann system with the use of the transformation~(\ref{horoz1})
after introduction of a gyrostat.

\subsection{A Generalization of the Kolosoff Analogy}

G.\,V.\,Kolosoff considered a transformation in~\cite{Kolosov1} which
 involves coordinates and time and reduces the Kovalevskaya
problem on~$e(3)$ to dynamics of a point on the Euclidean plane
in~$\mathbb{R}^2$ in a potential field, for which variables can be
separated. This transformation is used to introduce
action-angle variables in the Kovalevskaya problem in~\cite{KomKuz1}.

Let us consider an analogous procedure for the Kovalevskaya problem on the
bundle (1). In this case an analog of the Kolosoff transformation
gives dynamics of a particle on an axisymmetric nonconstant curvature surface.

Following~\cite{KomKuz1}, let us express the Hamiltonian from the motion
equations. Let us change coordinates in~(\ref{KOV10_1})
\begin{equation}
\label{m4.1}
s_i\to s_i-\frac {1}{2}(h+\frac {x}{4}),\quad i=1,\,2
\end{equation}
and represent them in the form
$$
 \frac{(s_1-s_2)^2 \dot s_i^2}{f(s_i)}=g(s_i),\quad i=1,\,2
$$
\begin{equation}
\label{m4.2}
\begin{aligned}
f(s) & = 4(s-\frac {x}{4}+\frac {1}{2} k)(s-\frac {x}{4}-\frac {1}{2}k),\\
g(s) & = 4s^3-(2h+\frac {3}{2} x)s^2+(c-k^2+\frac{x^2}{4})s+\varkappa,\\
\varkappa & = \frac {1}{2}(k^2h+2l^2-ch)+\frac {x}{4}(h^2+k^2-c)-\Bigl(\frac
{x}{4}\Bigr)^3.
\end{aligned}
\end{equation}

{\bf Remark.} {\it
The form of the coordinate change~(\ref{m4.1}) is dictated
by the requirement that the energy constant in~$g(s)$ was contained
in even powers of~$s$  only.}

Substraiting the first equation from the second in~(\ref{m4.2}), we eliminate
the constant $\varkappa$ and express the energy from the obtained relation
\begin{equation}
\label{m4.3}
\begin{aligned}
 H & = \frac{(s_1-s_2)^2}{2(s_1^2-s_2^2)}\Bigl(\frac{\dot s_1^2}{f(s_1)}-
 \frac{\dot s_2^2}{f(s_2)}\Bigr)+U(s_1,\,s_2),\\
 U(s_1,\,s_2) & = \frac{2(s_1^2+s_1s_2+s_2^2)+\frac 12(c-k^2+\frac{x^2}{4})}
 {s_1+s_2}.
\end{aligned}
\end{equation}

After the change of time~$d\tau=2(s_1+s_2)\,dt$ and transition to the
canonical momenta
$p_i=\frac{\partial {H}}{\partial{s_i'}}$, \mbox{$s_i'=\frac{ds_i}{d\tau},\;i=1,\,2$}, one obtains
a system with separable variables.

Let us consider the variables~$s_1-\frac {x}{4},\;s_2-\frac {x}{4}$
as elliptic coordinates on a plane with
the following relation with the  Cartesian coordinates ~$u,\,v$
$$
\begin{aligned}
 u & = \frac {2}{k}\Bigl(s_1-\frac {x}{4}\Bigr)\Bigl(s_2-\frac {x}{4}\Bigr)+\frac {k}{2},\\
 v & = \pm\frac {2}{k}\sqrt{\Bigl(s_1^2-\frac{k^2}{4}\Bigr)
 \Bigl(\frac{k^2}{4}-s_2^2\Bigr)}.
\end{aligned}
$$
We obtain a system on a plane with the energy determined by the relations
\begin{equation}
\label{m4.4}
\begin{aligned}
 H=&T+U,\\
 T=&\frac {1}{2}\Bigl(1+\frac{x}{2\rho}\Bigr)({u'}^2+{v'}^2),\\
 U=& \frac{4\rho^2-2uk+c+3x\rho+x^2}{2\rho+x}=
       \frac{2(\rho^2+\rho_1^2)-k^2+c+3x\rho+x^2}{2\rho+x},\\
 \rho=&\sqrt{u^2+v^2},\quad \rho_1=\sqrt{(u-k)^2+v^2}.
\end{aligned}
\end{equation}

The system~(\ref{m4.4}) describes the motion of a point along a curvilinear
surface in a potential field. The Gauss curvature can be evaluated by the
metrics which is defined by the kinetic energy $T$
$$
  K=-\frac{4x}{(2\rho+x)^3}.
$$

\subsection{The Action-Angle Variables for
the Kovalevskaya Top on the Bundle}

The method of introduction of the action-angle variables developed in this paper
is analogous to the method considered in~\cite{KomKuz1}. In comparison
with a usually cited procedure due to A.\,P.\,Veselov and S.\,P.\,Novikov,
this algorithm is more natural and uses an ordinary method of
introduction of action-angle variables for the systems with separating
variables.

Our algorithm consists of the following steps:

1. Construction of the Abel variables $s_1$, $s_2$, that commute
in the original Poisson structure: $\{s_1,\,s_2\}=0$.

{\bf Remark 1.}  {\it
The existence of commuting set of the Abel  variables   $s_1$, $s_2$
can be established with the help of arguments due to~\cite{KozlovMethods}.
These arguments are connected with the reduction of the equations~(\ref{KOV10_1})
to the standard form on a torus.}

2.
Introduction of the canonical momenta $p_i$ with the help of the energy
equation in the variables $s_i$, $\dot{s_i}$. These momenta must satisfy
an additional requirement that the system  $p_i$, $s_i$ should
be separable.

3. Having a set of separated variables, the action-angle variables can be
introduced in accordance with a well known algorithm.

It is easy to verify that the Kovalevskaya variables~(\ref{KOV9})
$s_1$, $s_2$ commute. They also satisfy the equations~(\ref{KOV10_1})
\begin{equation}
\label{act1}
\dot{ s_1}=\frac{\sqrt{f(s_1)\varphi(s_1)}}{s_1-s_2}, \quad
\dot{ s_2}=-\frac{\sqrt{f(s_2)\varphi(s_2)}}{s_1-s_2},
\end{equation}
where
\begin{equation}
\label{act2}
\begin{aligned}
f(s)={}&\left(2s+\frac{1}{4}(2h-x)+\frac{1}{4}k\right)
      \left(2s+\frac{1}{4}(2h-x)-\frac{1}{4}k\right),\\
\varphi(s) ={}& 4s^3 + 2hs^2
+\left(\frac{1}{16}(2h-x)^2-\frac{k^2}{16}+\frac{c}{4}\right) s+\frac{l^2}{16}.
\end{aligned}
\end{equation}
(The variable $s_1$ varies from  0 to $\infty$, and $s_2$
parameterizes a circle.)
\goodbreak

Let us extract a motion integral
$\varkappa=\frac{1}{16}\left((2h-x)^2-k^2\right)$ in~(\ref{act2}):
\begin{equation}
\label{act5}
\begin{aligned}
f(s)={}&4s^2+(2h-x)s+\varkappa,\\
\varphi(s)={}&s(4s^2+2hs+\varkappa+\frac{c}{4})+\frac{l^2}{16}\\
\end{aligned}
\end{equation}
and exclude  the variable $\varkappa$ from~(\ref{act5}). The obtained
equation express the energy  $h$ as a function \mbox{of $s_i$ and  $\dot{s_i}$}:
\begin{equation}
\label{act6}
\begin{array}{c}
h=-2(s_1+s_2)+\frac{l^2}{64s_1s_2}+\frac{x}{4}+
\frac{\sqrt{a_1^2+x_1^2}-\sqrt{a_2^2+x_2^2}}{s_1-s_2},\\
a_i=\frac{16 s_i^2x+4 s_ic+l^2}{64s_i},
\quad x_i=\frac{(s_1-s_2)\dot{s_i}}{2\sqrt{s_i}}.
\end{array}
\end{equation}

Let us introduce the conjugated momenta $p_i$  instead of the velocities $\dot{s_i}$:
\begin{equation}
\label{act61}
p_i=\int \frac{\partial h}{\partial \dot{s_i}} \frac{d\dot{s_i}}{\dot{s_i}}+F(s_i),
\end{equation}
where $F(s_i)$ is an arbitrary function of $s_i$. Note that the
  addition of
 $F(s_i)$ does not change the motion equations (the canonical
transformation).

We obtain
\begin{equation}
\label{act7}
p_i=\frac{1}{2\sqrt{s_i}}\ln \frac{x_i+\sqrt{x_i^2+a_i^2}}{a_i}.
\end{equation}

Let us write down the Hamiltonian of the Kovalevskaya system~(\ref{act6})
in the variables $s_i$, $p_i$
\begin{equation}
\label{act8}
h=-s_1-s_2+\frac{l^2}{8s_1s_2}+\frac{x}{8}+
\frac{a_1\cosh(2p_1\sqrt{s_1})-a_2\cos(2p_2\sqrt{-s_2})}{s_1-s_2}.
\end{equation}

The variables $s_i$ are separating for the Hamiltonian~(\ref{act8}). Introducing
the constant of separation~$\varkappa_1$, we obtain two equations,
which can be integrated separately from each other:
\begin{equation}
\label{act9}
\begin{array}{l}
2s_1^2+s_1\left(h-\frac{x}{4}\right)+\frac{l^2}{64s_1}+\varkappa_1={a_1}\cosh(2p_1\sqrt{s_1}),\\
2s_2^2+s_2\left(h-\frac{x}{4}\right)+\frac{l^2}{64s_2}+\varkappa_1={a_2}\cos(2p_2\sqrt{-s_2}).\\
\end{array}
\end{equation}

Inserting  (\ref{act7})  in~(\ref{act9}), we obtain
the Kovalevskaya equations~(\ref{act1}). The separation constant~$\varkappa_1$
is connected with the constant~$\varkappa$  by the formula
 $\varkappa=2\varkappa_1-\frac{c}{8}$.

Thus, the action in the variables $s$ equals
$$
I=\frac{1}{2\pi}\oint p(s)\, ds,
$$
i.\,e. is proportional to the area, restricted by a phase curve. The integration
region depends on the parameters and values of integrals of the system.

\section{The Goryachev-Chaplygin Case (1903)}

\subsection{Integrals and Separation of Variables}

The body is dynamically symmetric in this case: \mbox{$I_1=I_2=4I_3$},
the center of mass is situated in the equatorial plane of the inertia ellipsoid.
The Hamiltonian and the additional integral have the form
\begin{equation}
\label{}
\begin{aligned}
H={}&\frac{1}{2}(M_1^2+M_2^2+4M_3^2)+\gamma_1,\\
F_3={}&M_3(M_1^2+M_2^2)+M_1\boldsymbol\gamma_3.
\end{aligned}
\end{equation}

The gyrostat in the Goryachev-Chaplygin case was introduced by L.\,N.\,Sretensky (1963):
$$
\begin{aligned}
H={}&\frac 12\biggl(M_1^2+M_2^2+4\Bigl(M_3-\frac k4\Bigr)^2\biggr)+r_1\gamma_1,\\
F={}&(2M_3-k)(M_1^2+M_2^2)-2r_1M_1\gamma_3.
\end{aligned}
$$

Let us note that the gyrostatic moment is directed along the axis of
dynamical symmetry in generalizations of the Kovalevskaya and
Goryachev-Chaplygin cases. A separation of variables in the
Sretensky case (a generalization of the Goryachev-Chaplygin case)
is described in~\cite{sret}.

A generalization of the Goryachev-Chaplygin case
at the zero area constant was suggested in the work of
D.\,N.\,Goryachev~\cite{goriachev1}. The generalized Hamiltonian and
the integral have the form
\begin{equation}
\label{gen1}
\begin{gathered}
\begin{aligned}
H&=\frac{1}{2}(M_1^2+M_2^2+4M_3^2)-r\gamma_1+\frac{a}{\gamma_3^2},\\
F&=M_3\left(M_1^2+M_2^2+2\frac{a}{\gamma_3^2}\right)+r\gamma_3M_1,\\
\end{aligned}\\
a,\;r=const.
\end{gathered}
\end{equation}

 I.\,V.\,Komarov and V.\,B.\,Kuznetsov in~\cite{KomKuz}
added a constant gyrostatic moment to this Hamiltonian
(also at $(\boldsymbol M,\,\boldsymbol\gamma)=0$) analogous to the
Sretensky generalization
\begin{equation}
\label{gen2}
\begin{gathered}
\begin{aligned}
H&=\frac{1}{2}(M_1^2+M_2^2+4M_3^2)-r\gamma_1+\lambda M_3+\frac{a}{\gamma_3^2},\\
F&=\left(M_3+\frac{\lambda}{2}\right)\left(
M_1^2+M_2^2+2\frac{a}{\gamma_3^2}\right)+r\gamma_3M_1,\\
\end{aligned}\\
a,\; r,\; \lambda=const.
\end{gathered}
\end{equation}

%%%%%%%%%%%%%%%%%%%%%%%%%%%%%%%%%%%%%%%%%%%%%%%%%%%%%%%%
We shall give an analog of the Andoyer-Deprit variables for the bundle
of brackets $\mathcal{L}_x$. Let us use of the following sequent of
embeddings
$\mathbb R^1\subset SO(3)\subset \mathcal{L}_x$.

Let~$M_3$ be equal to one of the momentum variables
\begin{equation}
\label{m6.1}
M_3=L.
\end{equation}
Its canonically conjugated coordinate~$l(\{l,\,L\}=1)$ on~$so(3)$
can be found by integration of the Hamiltonian flow with the Hamiltonian~$\mathcal H=L$
\begin{equation}
\label{m6.2}
\begin{aligned}
 \frac{dM_1}{dl}  =\{M_1,\,L\}=M_2, \quad
 \frac{dM_2}{dl}  =\{M_2,\,L\}=-M_1,\quad
 \frac{dM_3}{d\tau}  =\{M_3,\,L\}=0.
\end{aligned}
\end{equation}
Further taking into account the commutation relation~$\{M_2,\,M_2\}=-M_3$
we obtain
\begin{equation}
\label{m6.3}
M_1=\sqrt{G^2-L^2}\sin l,\quad M_2=\sqrt{G^2-L^2}\cos l,
\end{equation}
where~$G^2=M_1^2+M_2^2+M_3^2$~is a Casimir function of the subalgebra~$so(3)$.
Let us choose~$G$ as the other momentum variable. Then the corresponding
flow on the total algebra~$\mathcal{L}_x$ has the form
\begin{equation}
\label{m6.4}
\frac{d\boldsymbol M}{dg}=0,\quad \frac{d\boldsymbol
p}{dg}=\frac 1G \boldsymbol p\times \boldsymbol M,
\end{equation}
where~$g$ is the variable canonically conjugated to~$G$.

{\bf Remark.} {\it The choice of~$G$ as a new canonical variable (not~$G^2$)
is determined by the fact that the corresponding variable~$g$
varies in the range from~0 to~$2\pi$ and does not depend on the value
of~$G$.}

Accordingly to~(\ref{m6.4}), $\boldsymbol M$ does not depend on~$g$.
Using~(\ref{m6.4}) for $\boldsymbol p$ and the Casimir functions
$$
 (\boldsymbol M,\,\boldsymbol p)=H,\quad p^2+xM^2=c
$$
we obtain
\begin{equation}
\label{m6.5}
\begin{aligned}
\boldsymbol p=\frac{H}{G^2}\boldsymbol M+\frac{\alpha}{G}(\boldsymbol M\times \boldsymbol e_3\sin g+G\boldsymbol
M\times(\boldsymbol M\times \boldsymbol e_3)\cos g),\\
\alpha^2=\frac{c-xG^2-\frac{H^2}{G^2}}{G^2-L^2},\quad \boldsymbol e_3=(0,\,0,\,1).
\end{aligned}
\end{equation}

Thus,~(\ref{m6.1}),~(\ref{m6.3}),~(\ref{m6.5}) define symplectic
coordinates on the entire bundle~$\mathcal{L}_x$. These coordinates at~$x=0$
pass into known Andoyer-Deprit variables in rigid body dynamics.

We shall find a generalization of the Goryachev-Chaplygin case
on $\mathcal{L}_x$ using~(\ref{m6.1}),~(\ref{m6.3}),~(\ref{m6.5}).
Let us take the Hamiltonian in the form
\begin{equation}
\label{m6.7}
\mathcal{H}=\frac {1}{2}(G^2+3L^2)+\lambda L+a\bigr(\cos l\cos g+\frac LG\sin l\sin g\bigl),
\end{equation}
where~$a,\,\lambda$~ are constants.

In comparison with~\cite{KozlovMethods},
a linear with respect to $L$ term describing the gyrostat is added in
(\ref{m6.7}).

The system~(\ref{m6.7}) admits a separation of variables. Indeed,
let us perform a canonical change of variables
\begin{equation}
\label{Dop*}
 L=p_1+p_2, \quad G=p_1-p_2, \quad q_1=l+g,\quad q_2=l-g.
\end{equation}
The Hamiltonian~(\ref{m6.7}) can be represented as
\begin{equation}
\label{m6.8}
\mathcal{H}=\frac {1}{2}\frac{p_1^3-p_2^3}{p_1-p_2}-\lambda\frac{p_1^2-p_2^2}{p_1-p_2}+
\frac{a}{p_1-p_2}(p_1\sin q_1+p_2\sin q_2).
\end{equation}
Let us express the Hamiltonian~(\ref{m6.7}) via the variables~$\boldsymbol M,\,\boldsymbol p$
at the zero area constant
$$
(\boldsymbol M,\,\boldsymbol p)=H=0.
$$
We obtain
$$
 \mathcal{H}=\frac {1}{2}(M_1^2+M_2^2+4M_3^2)-\lambda M_3+a\frac{p_1}{|\boldsymbol p|}.
$$

The additional integral in this case has the form
$$
 \mathcal{H}=(M_3-\frac{\lambda}{2})(M_1^2+M_2^2)-a M_1\frac{p_3}{|\boldsymbol p|}.
$$

For the algebra~$e(3)\;|\boldsymbol p|=1$, thus we obtain the Goryachev-Chaplygin case.

\subsection{The Action-Angle Variables for the Goryachev-Chaplygin Case}

The variables $q_1$, $q_2$~(\ref{Dop*}) are already separable. Moreover,
they commute with each other as it is easy to verify.

Let  $\varkappa$ denote the separation constant. Then
\begin{equation}
\label{gorchap1}
\begin{aligned}
\frac{p_1^3}{2}-\lambda p_1^2+ap_1\sin q_1 -hp_1={}&\varkappa,\\
\frac{p_2^3}{2}-\lambda p_2^2-ap_2\sin q_2 -hp_2={}&\varkappa.
\end{aligned}
\end{equation}

{\bf Remark 2.} {\it
Since}
\begin{equation}
\label{gorchap2}
\dot{p_i}=-\frac{\partial H}{\partial q_i}=-\frac{ap_i\cos q_i}{p_1-p_2},
\end{equation}
{\it using (\ref{gorchap1}) we obtain the Abel equations in the form}
\begin{equation}
\label{gorchap3}
\begin{aligned}
\dot{p_i}&=-\frac{\sqrt{\Phi(p_i)}}{2(p_1-p_2)},\\
\Phi(z)&=4a^2z^2-(2\varkappa-z^3+2\lambda z^2+2hz)^2.
\end{aligned}
\end{equation}

The action in the variables $s$ equals
$$
I=\frac{1}{2\pi}\oint p(s)\, ds.
$$

%%%%%%%%%%%%%%%%%%%%%%%%%%%%%%%%%%%%%%%%%%%%%%%%%%%%%%%%%%%%%%%%
\section{The Chaplygin Case}

\subsection{The Integrals and the Separation of Variables}

This case discovered by Chaplygin in 1902~\cite{chapl}
is a particular integrable case at the zero area constant
 $(\boldsymbol M,\boldsymbol \gamma)=0$ and a quartic integral. This system is
similar to the Kovalevskaya case in the Euler-Poisson equations.

The Hamiltonian and the additional integral of the system have the form~\cite{chapl}
\begin{equation}
\label{q}
\begin{aligned}
H={}&\frac{a}{2}(M_1^2+M_2^2+2M_3^2)+\frac{c}{2}(\gamma_1^2-\gamma_2^2), \\
F={}&(M_1^2-M_2^2+\frac {c}{a}\boldsymbol\gamma_3^2)^2+4M_1^2M_2^2.
\end{aligned}
\end{equation}

Let us consider the integration of the Kirchhoff equations using
the method of separation of variables  in the Chaplygin case
on the bundle of brackets $\mathcal{L}_x$
(the integration in this case has not been indicated before).
The separation of variables in this problem on  was performed by
S.\,A.\,Chaplygin~\cite{chapl} $e(3)$,
and by O.\,I.\,Bogoyavlensky~\cite{Bogoyav} on $so(4)$.

The Hamiltonian and the integrals on the bundle have the form
$$
\begin{array}{c}
H=\frac{1}{2}(\alpha_2 M_1^2+\alpha_1 M_2^2+(\alpha_1+\alpha_2) M_3^2
-(a_1-a_2)(\gamma_1^2-\gamma_2^2))=\frac{h}{2},\\[2mm]
J_2=x(M_1^2+M_2^2+M_3^2)+\gamma_1^2+\gamma_2^2+\gamma_3^2,\\[2mm]
J_3=M_1\gamma_1+M_2\gamma_2+M_3\gamma_3=0,\\[2mm]
J_4=(\alpha_1 M_1^2-\alpha_2 M_2^2-(a_1-a_2)\gamma_3^2)^2+
 4\alpha_1 \alpha_2 M_1^2M_2^2=k^2,
\end{array}
$$
where $\alpha_1=1-xa_1$, $\alpha_2=1-xa_2$ ($k>0$).
Let us introduce the separating variables
$s_1$ and $s_2$ by formulae~\cite{Bogoyav, chapl}
\vspace{-2mm}
\begin{equation}
\label{abel0}
\begin{array}{c}
s_1=\frac{u+k}{v}, \quad s_2=\frac{u-k}{v}, \\[2mm]
u=\alpha_1 M_1^2+\alpha_2 M_2^2, \quad v=(a_2-a_1)\gamma_3^2.
\end{array}
\end{equation}

The evolution of $s_1$ and $s_2$ is determined by degenerate Abel-Jacobi
equations:
\begin{equation}
\label{abel1}
\dot s_1 = -\sqrt{(1-s_1^2)(\delta_1-\beta_1 s_1)}, \quad
\dot s_2 = -\sqrt{(1-s_2^2)(\delta_2-\beta_2 s_2)},
\end{equation}
where
$$
\begin{array}{c}
\delta_1=m_1(h+k)-m_2J_2, \quad
\delta_2=m_1(h-k)-m_2J_2,\\[2mm]
\beta_1=x(a_1-a_2)(h+k)-m_3J_2, \quad
\beta_2=x(a_1-a_2)(h-k)-m_3J_2,\\[2mm]
m_1=\alpha_1+\alpha_2, \quad
m_2=x(a_1-a_2)^2,        \quad
m_3=(a_1-a_2)(\alpha_1+\alpha_2).
\end{array}
$$
Thus, the motion equations can be integrated by means of elliptic functions of time.

\subsection{The Action-Angle Variables for the Chaplygin System on the Bundle\\
of Brackets}

The variables $s_1$, $s_2$~(\ref{abel0}) commute at the zero
value of the integral  $(\boldsymbol M,\,\boldsymbol\gamma)=0$.

We find the energy $E=2h$ as a function of $s_1$, $s_2$, $\dot{s_1}$, $\dot{s_2}$
from~(\ref{abel1}):
\begin{equation}
\label{chact1}
\begin{aligned}E={}&\frac{\dot{s_1}^2}{(1-s_1^2)(s_1 a_{12}+m_1)} +
\frac{\dot{s_2}^2}{(1-s_2^2)(s_2 a_{12}+m_1)} -\frac{2J_2 m_3}{a_{12}}+ \\
{}&+\frac{J_2(m_2a_{12}+m_1m_3)}{a_{12}}\left(
\frac{1}{s_1a_{12}+m_1}+ \frac{1}{s_2a_{12}+m_1} \right),
\end{aligned}
\end{equation}
where $a_{12}=x(a_2-a_1)$.
Introducing the conjugated momenta~(\ref{act61})
$$
p_i=\frac{2 \dot{s_i} }{ (1-s_i^2)(s_i a_{12}+m_1) },
$$
we obtain the Hamiltonian in separated variables
\begin{equation}\label{chact2}
\begin{aligned}
H={}&\frac{1}{4}(1-s_1^2)(s_1a_{12}+m_1)p_1^2
+\frac{1}{4}(1-s_2^2)(s_2a_{12}+m_1)p_2^2
-\frac{2J_2 m_3}{a_{12}}+ \\
{}&+\frac{J_2(m_2a_{12}+m_1m_3)}{a_{12}} \left(
\frac{1}{s_1 a_{12} +m_1}+ \frac{1}{s_2 a_{12} +m_1} \right).
\end{aligned}
\end{equation}

As it may be expected, the variables are separated in the Hamiltonian~(\ref{chact2}):
\begin{equation}\label{chact3}\begin{aligned}
\frac{1}{4}(1-s_1^2)(s_1a_{12}+m_1)p_1^2 +
\frac{J_2(m_2a_{12}+m_1m_3)}{a_{12}(s_1 a_{12} +m_1)}={}&\varkappa,\\
\frac{1}{4}(1-s_2^2)(s_2a_{12}+m_1)p_2^2
+\frac{J_2(m_2a_{12}+m_1m_3)}{a_{12}(s_2 a_{12} +m_1)}
={}&h+\frac{2J_2 m_3}{a_{12}}-\varkappa.
\end{aligned}
\end{equation}

The action in  $s$ equals
$$
I=\frac{1}{2\pi}\oint p(s)\, ds.
$$

\section{The Bogoyavlensky System}

The particular Bogoyavlensky case~\cite{Bogoyav} with the Hamiltonian
$$
H=\frac{1}{2}\left(a_1M_1^2+a_2M_2^2+a_3M_3^2+\frac{1}{2x}((a_2+a_3)\gamma_1^2
+(a_1+a_3)\gamma_2^2+(a_1+a_2)\gamma_3^2)\right)=\frac{h}{2}
$$
can be integrated by means of elliptic functions. The system possesses the integrals
$$
\begin{array}{l}
J_2=x(M_1^2+M_2^2+M_3^2)+\gamma_1^2+\gamma_2^2+\gamma_3^2,\\[2mm]
J_3=M_1\gamma_1+M_2\gamma_2+M_3\gamma_3=0,\\[2mm]
J_4=((a_3-a_2)\gamma_1^2+(a_1-a_3)\gamma_2^2+(a_1-a_2)\gamma_3^2)^2
+4(a_3-a_1)(a_3-a_2)\gamma_1^2\gamma_2^2=h^2;
\end{array}
$$
On $so(4)$ the integration method is developed in~\cite{Bogoyav}.
If we introduce the variables
$$
s_1=\frac{u+h}{v}, \quad s_2=\frac{u-h}{v},
$$
where $u=(a_3-a_2)\gamma_1^2+(a_3-a_1)\gamma_2^2$, $v=(a_1-a_2)\gamma_3^2$,
the motion equations can be reduced to the form
\begin{equation}
\label{abel2}
\dot s_1 = -\sqrt{(1-s_1^2)(\delta_1-\beta_1 s_1)/2}, \quad
\dot s_2 = -\sqrt{(1-s_2^2)(\delta_2-\beta_2 s_2)/2},
\end{equation}
and the constants $\delta_i$, $\beta_i$ are expressed as follows:
$$
\begin{array}{c}
\delta_1=(J_1+h/2)n_1+J_2n_2,\quad
\delta_2=(J_1-h/2)n_1+J_2n_2,\\[2mm]
\beta_1=(a_1-a_2)(J_2a_3-J_1-h/2), \quad
\beta_2=(a_1-a_2)(J_2a_3-J_1+h/2),\\[2mm]
 n_1=2a_3-a_1-a_2,\quad n_2=2a_1a_2-a_3(a_1+a_2).
\end{array}
$$
{\bf Remark 3.} {\it
Let us note that~(\ref{abel1}, \ref{abel2}) are degenerate
Abel-Jacobi equations, i.\,e. each of them depends on a unique
variable  $s_1$ or $s_2$ only, and a two-dimensional Abel torus
splits into one-dimensional tori.}

\section{Integrable Systems on a Sphere with a Cubic Integral\\
(the Gaffet System)}

In the B.\,Gaffet work~\cite{gaffet1} an integrable case on the two-dimensional
sphere $\gamma_1^2+\gamma_2^2+\gamma_3^2=1$ with the potential
$U=a (\gamma_1\gamma_2\gamma_3)^{-2/3}$ is found.

This integrable system on $e(3)$ and at the zero value of the area integral
possesses the integrals
\begin{equation}
\label{tsig1}
\begin{aligned}
H={}&\frac{1}{2}(M_1^2+M_2^2+M_3^2)-
   \frac{1}{2}\frac{a^2(\gamma_1^2+\gamma_2^2+\gamma_3^2)}
{(\gamma_1\gamma_2\gamma_3)^{2/3}}, \\
J={}&M_1M_2M_3+a^2\left(\frac{M_1}{\gamma_1}+\frac{M_2}{\gamma_2}+
\frac{M_3}{\gamma_3}\right)
    (\gamma_1\gamma_2\gamma_3)^{1/3}.
\end{aligned}
\end{equation}

Let us note that the integrability of the system and the form of integrals
(\ref{tsig1}) are preserved on the total bundle of brackets $\mathcal{L}_x$
and at the zero value $(\boldsymbol M,\,\boldsymbol\gamma)=0$.

Despite some particular results~\cite{gaffet, gaffet1}, an explicit
integration of~(\ref{tsig1}) has not been performed yet. $L-A$ pair
for this system is indicated in~\cite{tsig} and has the form:
$$
\begin{array}{c}
\frac{d}{dt}{\bf L}=[{\bf L},\,{\bf A}],\\[3mm]
{\bf L}=\left(
\begin{array}{ccc}
\lambda & M_3+ ay_3 & M_2-ay_2 \\
 M_3- ay_3 & \lambda & M_1+ay_1\\
 M_2+ ay_2 &  M_1-ay_1 & \lambda
\end{array}
\right),\\[7mm]
{\bf A}=\frac{2a}{3}(\boldsymbol y,\,\boldsymbol y)
\left(
\begin{array}{ccc}
0 & {y_3^{-1}} & {y_2^{-1}} \\
 {y_3^{-1}} & 0 & {y_1^{-1}} \\
{y_2^{-1}} &  {y_1^{-1}} & 0
\end{array}
\right),
\end{array}
$$
where $\boldsymbol y=\frac{\boldsymbol \gamma}{(\gamma_1\gamma_2\gamma_3)^{1/3}}$.

\end{document}